\documentclass[showkeys,showpacs,superscriptaddress]{revtex4}
\usepackage{amsmath}
\usepackage{amssymb}
\usepackage{graphicx}
\usepackage{float}
\usepackage{xcolor}
\allowdisplaybreaks
\DeclareMathOperator\arctanh{arctanh}

\begin{document}

\title{
Nonperturbative quantization approach for QED on the Hopf bundle
}

\author{
Vladimir Dzhunushaliev
}
\email{v.dzhunushaliev@gmail.com}
\affiliation{
Department of Theoretical and Nuclear Physics,  Al-Farabi Kazakh National University, Almaty 050040, Kazakhstan
}
\affiliation{
Institute of Experimental and Theoretical Physics,  Al-Farabi Kazakh National University, Almaty 050040, Kazakhstan
}
\affiliation{
National Nanotechnology Laboratory of Open Type,  Al-Farabi Kazakh National University, Almaty 050040, Kazakhstan
}

\affiliation{
Academician J.~Jeenbaev Institute of Physics of the NAS of the Kyrgyz Republic, 265 a, Chui Street, Bishkek 720071, Kyrgyzstan
}

\author{Vladimir Folomeev}
\email{vfolomeev@mail.ru}
\affiliation{
Institute of Experimental and Theoretical Physics,  Al-Farabi Kazakh National University, Almaty 050040, Kazakhstan
}
\affiliation{
National Nanotechnology Laboratory of Open Type,  Al-Farabi Kazakh National University, Almaty 050040, Kazakhstan
}

\affiliation{
Academician J.~Jeenbaev Institute of Physics of the NAS of the Kyrgyz Republic, 265 a, Chui Street, Bishkek 720071, Kyrgyzstan
}

\affiliation{
International Laboratory for Theoretical Cosmology, Tomsk State University of Control Systems and Radioelectronics (TUSUR),
Tomsk 634050, Russia
}

\begin{abstract}
We consider the Dirac equation and Maxwell's electrodynamics in $\mathbb{R} \times S^3$ spacetime, where a three-dimensional sphere is the Hopf bundle
$S^3 \rightarrow S^2$. In both cases, discrete spectra of classical solutions are obtained. Based on the solutions obtained,
the quantization of free, noninteracting Dirac and Maxwell fields is carried out.
The method of nonperturbative quantization of interacting Dirac and Maxwell fields is suggested.
The corresponding operator equations and the infinite set of the Schwinger-Dyson equations for Green's functions is written down.
To illustrate the suggested scheme of nonperturbative quantization, we write a simplified set of equations describing some physical situation.
Also, we discuss the properties of quantum states and operators of interacting fields.
\end{abstract}



\date{\today}

\maketitle

\section{Introduction}

Quantum electrodynamics (QED) and the electroweak theory are very successful in explaining quantum phenomena for electromagnetic and weak interactions. Their predictions agree with experimental data to great precision.
This progress was achieved despite the fact that the calculations are perturbative and one needs to involve, for example, a renormalization procedure. R.~Feynman called this procedure ``sweeping the garbage under the rug.'' L.~Landau  et al.~\cite{Landau:1} wrote on this subject:
``Although at present there exist methods to remove these singularities (regularization), which clearly lead to correct results, such method of action has the artificial nature. The singularities arise in the theory due to the pointlike interaction described by  delta functions (operators of interacting fields are taken at one point).'' In Refs.~\cite{Landau:2,Landau:3,Landau:4}, he and his co-authors study this question trying to remove such singularities in QED.

Attempts to use this technique for strong interactions do not lead to a full success, and in gravity they are unsuccessful. This suggests that at present we do not clearly understand the nature of quantization. It is reasonable to assume that there should exist some well-defined mathematical procedure of quantization which can be applied to \emph{any} field theory. This cannot be a perturbative technique used, for instance, in QED, since it leads to nonrenormalizable theories. This suggests that the aforementioned procedure of quantization, applicable to any field theory, must be nonperturbative. It should be pointed out here that the nonperturbative quantization technique was perhaps first suggested by  W.~Heisenberg in Ref.~\cite{heis} where he proposed an idea that an electron can be described by a nonperturbatively quantized nonlinear spinor field.

Taking all this into account, it is of great interest to construct nonperturbative QED for some simple case in order to compare perturbative and
nonperturbative QED. It would allow one to understand the physical essence of such phenomena like the renormalization, convergence of the Feynman integral, etc.
One can assume that such a possibility may arise in constructing QED on some compact manifold, since in this case spectra of eigenvalues
of the Dirac and Maxwell equations will presumably be discrete. This would permit one to replace the Fourier integral in a general solution by a summation over quantum numbers
that give the number of the eigenvalues. In doing so, Dirac delta functions would be replaced by the Kronecker symbols,
and all calculations would be simplified; this presumably would permit one to construct nonperturbative QED.
Simultaneously, it would be possible to construct perturbative QED according to conventional methods, but taking into account the compactness of three-dimensional space.
After that, there would appear the possibility of comparison of perturbative and nonperturbative quantum theories.

In the present paper, we find a discrete spectrum of classical solutions describing noninteracting Dirac and Maxwell fields in a spacetime with a spatial cross-section
in the form of the Hopf bundle. Then we use the spectra obtained to quantize the Dirac equation and Maxwell's electrodynamics.
 Finally, we suggest a scheme of nonperturbative quantization of coupled Dirac and Maxwell fields.

The paper is organized as follows. In Sec.~\ref{QED_D}, we give the Lagrangian, field equations, and $\mathfrak{Ansatze}$ for the Dirac and Maxwell
equations on the Hopf bundle. Using them, we obtain classical solutions to the Dirac (Sec.~\ref{Dirac_class})
and Maxwell (Sec.~\ref{Max_class}) equations separately.
In Sec.~\ref{quantum_lin_fields}, we quantize free, noninteracting fields and obtain expressions for the corresponding propagators.
 In Sec.~\ref{NP_QED}, we carry out the nonperturbative quantization on the Hopf bundle. Finally, in Sec.~\ref{discussion},
 we discuss the results obtained and list the important problems in nonperturbative quantum field theory.

\section{Classical electrodynamics plus the Dirac equation
}
\label{QED_D}

In this section we consider classical electrodynamics coupled to spinors obeying the Dirac equation in $\mathbb{R} \times S^3$
spacetime with a spatial cross-section in the form of the Hopf bundle $S^3 \rightarrow S^2$.
One can say that a relativistic quantum theory of an electron interacting with an electromagnetic field and living on the the Hopf bundle is under consideration.


Here we closely follow Ref.~\cite{Dzhunushaliev:2020pyk}. Consider Dirac-Maxwell theory with the source of electromagnetic field taken in the form of a massless Dirac field.
The corresponding Lagrangian can be chosen in the form (hereafter, we work in units where $\hbar=c=1$)
\begin{equation}
	L =	\frac{i} {2} \left(
			\bar \psi \gamma^\mu \psi_{; \mu} -
			\bar \psi_{; \mu} \gamma^\mu \psi
		\right) - \frac{1}{4} F_{\mu\nu}F^{\mu\nu}
\label{lagr_tot}
\end{equation}
with the covariant derivative
$
\psi_{; \mu} \equiv \left[\partial_{ \mu} +
1/8\, \omega_{a b \mu}\left(
	\gamma^a  \gamma^b -
	\gamma^b  \gamma^a\right) -
	i e A_\mu
\right]\psi
$,
where $\gamma^a$ are the Dirac matrices in
flat space (below we use the spinor representation of the matrices);
$a, b$ and $\mu, \nu$ are tetrad and spacetime indices, respectively;
$
	F_{\mu \nu} = \partial_\mu A_\nu - \partial_\nu A_\mu
$ is the electromagnetic field tensor; $A_\mu$ are four-potentials of the electromagnetic field; $e$ is a charge in Maxwell theory.
In turn, the Dirac matrices in curved space,
$\gamma^\mu = e_a^{\phantom{a} \mu} \gamma^a$, are derived using
the tetrad
$ e_a^{\phantom{a} \mu}$, and $\omega_{a b \mu}$ is the spin connection
[for its definition, see Ref.~\cite{Lawrie2002}, formula (7.135)].

Varying the corresponding action with the Lagrangian \eqref{lagr_tot}, one can derive the following set of equations:
\begin{align}
	i \gamma^\mu \psi_{;\mu} =& 0 ,
\label{D_10}\\
	\frac{1}{\sqrt{-g}}
	\frac{\partial }{\partial x^\nu}
	\left( \sqrt{-g} F^{\mu \nu} \right) =& - 4 \pi j^\mu,
\label{D_15}
\end{align}
where $g$ is the determinant of the metric tensor and
$j^\mu=e \bar\psi\gamma^\mu\psi$ is the four-current.

The above equations will be solved in $\mathbb{R} \times S^3$ spacetime with the Hopf coordinates $\chi, \theta, \varphi$ on a sphere~$S^3$ with
the metric
\begin{equation}
	ds^2 = dt^2 - \frac{r^2}{4}
	\left[
		\left( d \chi - \cos\theta d \varphi\right)^2
		+ d \theta^2 + \sin^2 \theta d \varphi^2
	\right] = dt^2 - r^2 dS^2_3 ,
\label{D_20}
\end{equation}
where $dS^2_3$ is the Hopf metric on the unit $S^3$ sphere; $r$ is a constant;
$0\leq \chi, \varphi \leq 2\pi$ and $0\leq \theta \leq \pi$.

To solve Eqs.~\eqref{D_10} and \eqref{D_15}, we employ the following
$\mathfrak{Ansatze}$ for the spinor and electromagnetic fields:
\begin{align}
	\psi_{nm} &= e^{-i  \Omega t} e^{i n \chi} e^{i m \varphi}
	\begin{pmatrix}
		\Theta_1(\theta)  \\
		\Theta_2(\theta)  \\
		0  \\
		0
	\end{pmatrix} ,
\label{D_30}\\
	A_\mu &= \left\lbrace
		\phi(\theta), r \rho(\theta) ,0, r \lambda(\theta)
	\right\rbrace  ,
\label{D_35}
\end{align}
where $m$ and $n$ are integers. The spinor can transform under a rotation through an angle $2 \pi$  as
$$
	\psi_{mn}(\chi + 2 \pi, \theta, \varphi + 2 \pi) =
	\psi_{mn}(\chi , \theta, \varphi).
$$
Then, because of the presence of the  factors $e^{i n \chi}$ and $e^{i m \varphi}$ in Eq.~\eqref{D_30},
the spinors $\psi_{mn}$ and $\psi_{pq}$ with different pairs of indices  $(m,n)$ and $(p, q)$ are  orthogonal.

To solve the equations, we use the tetrad
$$
  e^a_{\phantom{a} \mu} =
  \begin{pmatrix}
    1 & 		0 		& 0 			& 0 						\\
    0 & 	\frac{r}{2}	& 0 			& - \frac{r}{2} \cos \theta 	\\
    0 & 0				& \frac{r}{2} 	& 0 						\\
    0 & 0				& 0 			& \frac{r}{2} \sin \theta	
  \end{pmatrix}
$$
coming from the metric \eqref{D_20}.

On substitution of the $\mathfrak{Ansatze}$ \eqref{D_30} and \eqref{D_35} in Eqs.~\eqref{D_10} and \eqref{D_15},
we have
\begin{align}
	& \Theta_{1}^\prime + \Theta_{1} \left(
		\frac{\cot \theta}{2} + n + e r \rho
	\right) +
	\Theta_{2} \left(
		\frac{1}{4} - \frac{r \Omega}{2} - n \cot \theta -
		\frac{m}{\sin \theta} - e r \rho \cot \theta -
		\frac{e r \lambda}{\sin \theta}	+ \frac{e r}{2} \phi
	\right) = 0 ,
\label{D_50}\\
	&\Theta_{2}^\prime + \Theta_{2} \left(
		\frac{\cot \theta}{2} - n - e r \rho
	\right) + \Theta_{1}  \left(
		- \frac{1}{4} + \frac{r \Omega}{2} - n \cot \theta -
		\frac{m}{\sin	\theta}  - e r \rho \cot \theta -
				\frac{e r \lambda}{\sin \theta} -
				\frac{e r}{2} \phi
	\right) = 0 ,
\label{D_60}\\
	& \frac{1}{\sin \theta} \left(
		\sin \theta \phi^\prime
	\right)^\prime = \phi^{\prime \prime} +
	\cot \theta \phi^ \prime =
	- \frac{e r^2}{4} \left(
		\Theta_1^2 + \Theta_2^2
	\right) ,
\label{D_70}\\
	&\left(
		\frac{\rho^\prime}{\sin \theta}
	\right)^\prime +
	\left(
		\cot \theta \lambda^\prime
	\right)^\prime =
	\frac{\rho^{\prime \prime}}{\sin \theta} -
	\frac{\cot \theta}{\sin \theta} \rho^\prime +
	\cot \theta \lambda^{\prime \prime} -
	\frac{\lambda^\prime }{\sin^2 \theta} =
	\frac{e r^2}{8} \left[
		\cos \theta \left( \Theta_1^2 - \Theta_2^2 \right) +
		2 \sin \theta \Theta_1 \Theta_2
	\right] ,
\label{D_80}\\
	&\left(
		\frac{\lambda^\prime}{\sin \theta}
	\right)^\prime +
	\left(
		\cot \theta \rho^\prime
	\right)^\prime =
		\frac{\lambda^{\prime \prime}}{\sin \theta} -
		\frac{\cot \theta}{\sin \theta} \lambda^\prime +
		\cot \theta \rho^{\prime \prime} -
		\frac{\rho^\prime }{\sin^2 \theta} =
	\frac{e r^2}{8}
		\left( \Theta_1^2 - \Theta_2^2 \right) ,
\label{D_90}
\end{align}
where the prime denotes differentiation with respect to $\theta$. Notice that the set of equations
\eqref{D_50}-\eqref{D_90} must be solved as an eigenvalue problem for $\Omega$
with the eigenfunctions $\Theta_{1, 2}$.

\section{
Classical vacuum solutions to the Dirac equation
}
\label{Dirac_class}

In this section we consider the case of  ``frozen'' electric and magnetic fields with zero scalar and vector potentials, $\phi = \rho = \lambda = 0$.
In this case the parent Dirac equations~\eqref{D_50} and \eqref{D_60} take the form
\begin{align}
	\Theta_{1}^\prime + \Theta_{1} \left(
		\frac{\cot \theta}{2} + n
	\right) +
	\Theta_{2} \left(
		\tilde \Omega - n \cot \theta -
		\frac{m}{\sin \theta}
	\right) & = 0 ,
\label{g_sol_D_10}\\
	\Theta_{2}^\prime + \Theta_{2} \left(
		\frac{\cot \theta}{2} - n
	\right) + \Theta_{1}  \left(
		- \tilde \Omega - n \cot \theta -
		\frac{m}{\sin	\theta}
	\right) & = 0 ,
\label{g_sol_D_20}
\end{align}
where $\tilde \Omega = \frac{1}{4} - \frac{r \Omega}{2}$.
These equations are symmetric under the replacements
$
	\Theta_1 \rightarrow \Theta_2 ,
	\Theta_2 \rightarrow - \Theta_1 ,
	n \rightarrow - n , m \rightarrow - m
$.
Introducing new functions
$$
	\Sigma_1  = \frac{\Theta_{1} + \Theta_{2}}{2} , \quad
	\Sigma_2  = \frac{\Theta_{1} - \Theta_{2}}{2},
$$
Eqs.~\eqref{g_sol_D_10} and \eqref{g_sol_D_20} can be rewritten as
\begin{align}
	\Sigma_1^\prime + \Sigma_1 \left(
		\frac{1}{2} \cot \theta - n \cot \theta - \frac{m}{\sin \theta}
	\right) + \Sigma_2 \left( n - \tilde \Omega \right)
	& = 0 ,
\label{g_sol_D_50}\\
	\Sigma_2^\prime + \Sigma_2 \left(
		\frac{1}{2} \cot \theta + n \cot \theta + \frac{m}{\sin \theta}
	\right) + \Sigma_1 \left( n + \tilde \Omega \right)
	& = 0 ,
\label{g_sol_D_60}
\end{align}
which are in turn symmetric under the replacements
$
	\Sigma_2 \rightarrow - \Sigma_1 ,
	\Sigma_1 \rightarrow \Sigma_2 ,
	n \rightarrow - n , m \rightarrow - m .
$
Upon finding $\Sigma_2$ from Eq.~\eqref{g_sol_D_50},
\begin{equation}
	\Sigma_2 =- \frac{
		\Sigma_1^\prime + \Sigma_1 \left(
		\frac{1}{2} \cot \theta - n \cot \theta - \frac{m}{\sin \theta}
		\right)
	}{n - \tilde \Omega},
\label{g_sol_D_75}
\end{equation}
and substituting it in~\eqref{g_sol_D_60}, we get a second-order differential equation for the function  $\Sigma_1$,
\begin{equation}
	\Sigma_1'' + \cot \theta \Sigma_1' -
	\Sigma_1
	\frac{
		\frac{1}{4} + n^2 +m^2 - n +
		\left( 2 m n - m \right) \cos \theta
		+ \frac{1 - 4 \tilde \Omega^2}{4} \sin^2 \theta
	}{\sin^2 \theta} = 0 .
\label{g_sol_D_70}
\end{equation}
Similarly, for the function $\Sigma_2$, one can get the following equation:
\begin{equation}
	\Sigma_2'' + \cot \theta \Sigma_2' -
	\Sigma_2
	\frac{
		\frac{1}{4} + n^2 +m^2 + n +
		\left( 2 m n + m \right) \cos \theta
		+ \frac{1 - 4 \tilde \Omega^2}{4} \sin^2 \theta
	}{\sin^2 \theta} = 0.
\label{g_sol_D_80}
\end{equation}
It should be noted here that Eq.~\eqref{g_sol_D_80} can be obtained from Eq.~\eqref{g_sol_D_70} on simply replacing $m, n \rightarrow -m, -n$.

The equation~\eqref{g_sol_D_70} must be regarded as an eigenvalue problem for the parameter  $\tilde \Omega$.
This equation has the following general solution containing two linearly independent solutions:
\begin{equation}
\begin{split}
	\Sigma_1 = & c_1
	\left( 1 - \cos \theta \right)^{\alpha/2}
	\left( 1 + \cos \theta \right)^{\beta/2}
	{}_2 F_1 \left(
		n - \tilde \Omega, n + \tilde \Omega; n + m + \frac{1}{2};
		\frac{1 - \cos \theta}{2}
	\right)
\\
	&
	+c_2
	\left( 1 - \cos \theta \right)^{-\alpha/2}
	\left( 1 + \cos \theta \right)^{\beta/2}
		{}_2 F_1 \left(
			- m - \tilde \Omega + \frac{1}{2}, - m + \tilde \Omega + \frac{1}{2};
			- n - m + \frac{3}{2};
			\frac{1 - \cos \theta}{2}
		\right)
\label{g_sol_D_90}
\end{split}
\end{equation}
with
\begin{equation}
	\alpha = n + m -\frac{1}{2}, \beta = n - m -\frac{1}{2} .
\label{g_sol_D_95}
\end{equation}
Substituting \eqref{g_sol_D_90} in \eqref{g_sol_D_75}, we have
\begin{equation}
\begin{split}
	\Sigma_2 = & - c_1 \frac{n + \tilde \Omega}{2 n + 2 m + 1}
	\left( 1 - \cos \theta \right)^{(\alpha+1)/2}
	\left( 1 + \cos \theta \right)^{(\beta+1)/2}
	{}_2 F_1 \left(
		n + 1 - \tilde \Omega, n + 1 + \tilde \Omega;
		n + m + \frac{3}{2};
		\frac{1 - \cos \theta}{2}
	\right)
\\
	&
	+c_2 \left\lbrace
	\frac{(m - \frac{1}{2})^2 - \tilde \Omega^2}
	{(2 m + 2 n - 3) (n - \tilde \Omega )}
	\left( 1 - \cos \theta\right)^{(1-\alpha) /2}
	\left( 1 + \cos \theta\right)^{(\beta+1)/2}
	\right.
\\
	&
	\times {}_2F_1
	\left(
		- m - \tilde \Omega + \frac{3}{2},
		- m + \tilde \Omega + \frac{3}{2}; - m - n + \frac{5}{2};
		\frac{1 - \cos \theta}{2}
	\right)
\\
	&	
	+ \frac{n+m-1/2}{ (n - \tilde \Omega )}
	\left.
	\left( 1 - \cos \theta\right)^{- \alpha/2}
	\left( 1 + \cos \theta \right)^{1+\beta/2}
	\, _2F_1
	\left(
		- m - \tilde \Omega + \frac{1}{2},
		- m + \tilde \Omega + \frac{1}{2};
		- m - n + \frac{3}{2};
		\frac{1 - \cos \theta}{2}
	\right)
	\right\rbrace .
\label{g_sol_D_100}
\end{split}
\end{equation}
It is known that the hypergeometric function $_2F_1(a, b; c; x)$ is regular at the points $x = 0, 1$
if either  $a$ or $b$ is a negative integer. In our case this has the result that  for the first independent solution in~\eqref{g_sol_D_90} and \eqref{g_sol_D_100}
we have the quantization condition for the parameter~$\tilde \Omega$,
\begin{equation}
 \tilde \Omega_{n l} = \pm ( n + l ) , \text{ where }
 l = 0, 1, \ldots
\label{g_sol_D_110}
\end{equation}
In this case the hypergeometric function $_2F_1(a, b; c; x) = {}_2F_1(b, a; c; x)$ is equal to the Jacobi polynomials
\begin{equation}
 P^{\left( \gamma, \delta \right) }_l (x) =
 \frac{\left( \gamma + 1\right)_l}{l!} \;
 {}_2F_1 \left(
	-l, 1 + \gamma + \delta + l; \gamma + 1; \frac{1 - x}{2}
 \right),
\label{g_sol_D_120}
\end{equation}
where $\left( \gamma + 1\right)_l$ is the Pochhammer symbol. In our case this gives the following values of the parameters
$\gamma, \delta$ and the variable $x$:
$$
 x = \cos \theta, \gamma = \alpha = n + m - \frac{1}{2},
 \delta = \beta = n - m - \frac{1}{2} ,
 l = -n \mp \tilde \Omega .
$$
Thus, the first independent solution can be recast in the form
\begin{align}
	\Sigma_{1, n m l} &= c_1 \frac{l!}{\left( \alpha + 1\right)_l}
	\left( 1 - \cos \theta \right)^{\alpha/2}
	\left( 1 + \cos \theta \right)^{\beta/2}
	P^{\left( \alpha, \beta \right) }_{l} \left( \cos \theta\right) ,
\label{g_sol_D_140}\\
	\Sigma_{2, n m l} &= - c_1
	\frac{n + \tilde \Omega}{2 n + 2 m + 1}
	\frac{ l!}{\left( \alpha + 1\right)_l}
	\left( 1 - \cos \theta \right)^{(\alpha + 1)/2}
	\left( 1 + \cos \theta \right)^{(\beta + 1)/2}
	P^{\left( \alpha + 1, \beta + 1\right) }_{l - 1} \left( \cos \theta\right),
\label{g_sol_D_150}
\end{align}
where the parameters  $\alpha$ and $\beta$ are given by the expressions  \eqref{g_sol_D_95}.
In order to ensure the regularity of the factors $\left( 1 - \cos \theta \right)^{\alpha/2}$ and
$\left( 1 + \cos \theta \right)^{\beta/2}$, it is necessary that
 $\alpha \geqslant 0, \beta \geqslant 0$; this results in the following restrictions imposed upon the integers  $n$ and $m$:
$$
	n \geqslant 1 ,
	\left| m \right| \leqslant n - \frac{1}{2} .
$$
The functions $\Sigma_{1, n m p}$ with different $p$ and the functions $\Sigma_{2, n m p}$ are orthogonal:
\begin{align}
	\int \limits_{0}^{\pi} \sin \theta \;
	\Sigma_{1, n m p} \Sigma_{1, n m q} d \theta &=
	- c^2_1 \left[
		\frac{1}{\left( \alpha + 1\right)_p }
		\frac{2^{\alpha + \beta + 1}}{2 p + \alpha + \beta + 1}
		\frac{\Gamma(p + \alpha + 1) \Gamma(p + \beta + 1)}
		{\Gamma(p + \alpha + \beta + 1)}
	\right]^ 2 \delta_{pq}
	= A_{n m p} \delta_{pq} ,
\label{g_sol_D_160}\\
	\int \limits_{0}^{\pi} \sin \theta \;
	\Sigma_{2, n m p} \Sigma_{2, n m q} d \theta &=
	- c^2_1 \left[
		\frac{1}{\left( \alpha + 2\right)_p }
		\frac{2^{\alpha + \beta + 3}}{2 p + \alpha + \beta + 3}
		\frac{\Gamma(p + \alpha + 2) \Gamma(p + \beta + 2)}
		{\Gamma(p + \alpha + \beta + 3)}
	\right]^ 2 \delta_{pq}
	= B_{n m p} \delta_{pq}.
\label{g_sol_D_170}
\end{align}
Their orthogonality follows from the orthogonality condition for the Jacobi polynomials,
$$
	\int \limits_{-1}^{1}
	\left( 1 - x \right)^\rho \left( 1 + x \right)^\sigma
	P^{\left( \rho, \sigma \right) }_{p} \left( x \right)
	P^{\left( \rho, \sigma \right) }_{q} \left( x \right) dx =
	\delta_{p q} ,
$$
where $\rho=\beta, \sigma=\alpha$ for \eqref{g_sol_D_160} or  $\rho=(\beta+1), \sigma=(\alpha+1)$ for \eqref{g_sol_D_170} and
$x = -\cos \theta$. The functions $\Sigma_{1, n m p}$  and $\Sigma_{2, n m p}$ are not orthogonal to each other.

Using the solutions \eqref{g_sol_D_140} and \eqref{g_sol_D_150}, one can obtain the functions
\begin{align}
	\Theta_{1, nml} &=c_1
	\frac{l!}{\left( \alpha + 1\right)_l}
	\left( 1 - \cos \theta \right)^{\alpha/2}
	\left( 1 + \cos \theta \right)^{\beta/2}
	\left[
		P^{\left( \alpha, \beta \right) }_{l} \left( \cos \theta\right) -
		\frac{n + \tilde \Omega_{n l}}{2n+2m+1}\sin \theta
		P^{\left( \alpha + 1, \beta + 1\right) }_{l - 1} \left( \cos \theta\right)
	\right] ,
\label{g_sol_D_190}\\
	\Theta_{2, nml} &=
	c_1
	\frac{l!}{\left( \alpha + 1\right)_l}
	\left( 1 - \cos \theta \right)^{\alpha/2}
	\left( 1 + \cos \theta \right)^{\beta/2}
	\left[
		P^{\left( \alpha, \beta \right) }_{l} \left( \cos \theta\right) +
		\frac{n + \tilde \Omega_{n l}}{2n+2m+1} \sin \theta
		P^{\left( \alpha + 1, \beta + 1\right) }_{l - 1} \left( \cos \theta\right)
	\right]
\label{g_sol_D_200}
\end{align}
Consider the orthogonality of the corresponding spinors,
\begin{equation}
	\psi_{nml} = e^{-i  \Omega_{n l} t} e^{i n \chi} e^{i m \varphi}
	\begin{pmatrix}
		\Theta_{1, nml}  \\
		\Theta_{2, nml}
	\end{pmatrix}  =
	\psi_{+1/2, nml} + \psi_{-1/2, nml},
\label{g_sol_D_210}
\end{equation}
where $\psi_{\pm 1/2, nml}$ describe the states of particles with the projection of the spin $\pm 1/2$:
\begin{align}
	\psi_{+1/2, nml} = & \tilde c_1
	e^{-i  \Omega_{n l} t} e^{i n \chi} e^{i m \varphi}
	\left( 1 - \cos \theta \right)^{\alpha/2}
	\left( 1 + \cos \theta \right)^{\beta/2}
	P^{\left( \alpha, \beta \right) }_{l} \left( \cos \theta\right)
	\begin{pmatrix}
		1  \\
		1
	\end{pmatrix} ,
\label{spin_projection_plus}\\
	\psi_{-1/2, nml} = & \tilde c_1
	\frac{n + \tilde \Omega_{n l}}{2n+2m+1}
	e^{-i  \Omega_{n l} t} e^{i n \chi} e^{i m \varphi}
	\left( 1 - \cos \theta \right)^{(\alpha + 1)/2}
	\left( 1 + \cos \theta \right)^{(\beta + 1)/2}
	P^{\left( \alpha + 1, \beta + 1\right) }_{l - 1} \left( \cos \theta\right)
	\begin{pmatrix}
		- 1  \\
		1
	\end{pmatrix} .
\label{spin_projection_minus}
\end{align}
According to the remarks made after Eqs.~\eqref{g_sol_D_20} and \eqref{g_sol_D_60}, the following spinor is also the solution:
\begin{equation}
	\psi_{-n,-m,l} = e^{-i \Omega_{-n, l} t} e^{-i n \chi} e^{-i m \varphi}
	\begin{pmatrix}
		\Theta_{2; -n,-m,l}  \\
		\Theta_{1; -n, -m, l}
	\end{pmatrix} .
\label{add_soln}
\end{equation}
Hereafter, we drop for brevity the zeroth components of the spinor, i.e., we consider Weyl spinors. The orthogonality condition is
\begin{equation}
\begin{split}
	\left\langle \psi_{nml}, \psi_{pqr} \right\rangle = &
	\int \sqrt{\gamma} \;
		\psi_{nml}^\dagger \psi_{pqr} dV =
	4 \pi^2 \delta_{n p} \delta_{m q}
	\int \limits_0^\pi \sin \theta
	\left(
		\Theta_{1, nml} \Theta_{1, pqr} +
		\Theta_{2, pqr} \Theta_{2, pqr}
	\right) d \theta \nonumber
\\
	&
	=8 \pi^2
	\left(
		A_{n m l} + B_{n m l}
	\right) \delta_{n p} \delta_{m q} \delta_{l q} = 1 ,\nonumber
\end{split}
\end{equation}
where $dV = \sin \theta d \theta d \chi d \varphi$ and we took into account the orthogonality both of the functions
 $e^{i n \chi}$ with different integers $n$ and of the functions $e^{i m \chi}$ with different integers $m$;
 the normalization constant  $c_1$ is chosen from this condition using the values of
 $A_{n m l}$ and $B_{n m l}$ from \eqref{g_sol_D_160} and \eqref{g_sol_D_170}.

Consider the second independent solution in \eqref{g_sol_D_90}. It is regular when
$$
	\tilde \Omega = \pm \left[
	\left( m - \frac{1}{2}\right) - l
	\right] ,
$$
where $l$ is a nonnegative integer. Just as for the first independent solution, one can introduce the Jacobi polynomials according to~\eqref{g_sol_D_120}.
Then the solution takes the form
\begin{equation}
	\Sigma_1 = \frac{l!}{\left( \gamma + 1\right)_l}
		\left( 1 - \cos \theta \right)^{\gamma/2}
		\left( 1 + \cos \theta \right)^{\delta/2}
		P^{\left( \gamma, \delta \right) }_{l} \left( \cos \theta\right) ,
\label{g_sol_D_240}
\end{equation}
where the parameters $l, \gamma$, and $\delta$ are defined as follows:
$$
	l = \pm \left( m - \frac{1}{2}\right) - \tilde \Omega ,
	\gamma = -n -m + \frac{1}{2},
	\delta = n - m - \frac{1}{2} .
$$
In order to ensure the regularity of the factors $\left( 1 - \cos \theta \right)^{\gamma/2}$ and $\left( 1 + \cos \theta \right)^{\delta/2}$,
it is necessary that
$\gamma \geqslant 0, \delta \geqslant 0$;
this results in the following restrictions imposed upon the integers  $n$ and $m$:
$$
	m \leqslant 0 ,	n \leqslant \left| m \right| + \frac{1}{2} .
$$
Thus, the second independent solution \eqref{g_sol_D_240} has been reduced to the form \eqref{g_sol_D_140}
for the first independent solution; therefore, subsequent calculations are similar to those performed for the first independent solution, and we do not show them here.

Thus, the spinors \eqref{g_sol_D_210} form the complete basis set on a three-dimensional sphere in the Hopf coordinates.
It is important that this basis is numbered by integers, in contrast to Minkowski space where the complete basis set is created by plain waves and is numbered by real numbers.
This essential distinction follows directly from the fact that a three-dimensional sphere  $S^3$ is a compact space, in contrast to Minkowski space, which in turn is a noncompact space.

\section{Classical vacuum solutions to the Maxwell equations
}

\label{Max_class}

In this section we consider classical solutions with the ``frozen'' spinor field $\psi = 0$.
The four-potential  $A_\mu$ for the Maxwell equations~\eqref{D_15} can be written in the form
\begin{equation}
	A_\mu = e^{i \left(
		\frac{\omega}{r} t + p \chi + q \varphi
	\right) } \left\lbrace
		\phi(\theta), r \rho(\theta), 0, r \lambda(\theta)
	\right\rbrace,
\label{first_ans}
\end{equation}
where we have chosen the gauge $A_\theta = 0$, and $p, q$ are integers.  Using this potential, the Maxwell equations yield
\begin{align}
	\phi '' + \cot \theta \phi ' -
	\frac{p^2 + q^2 + 2 p q \cos \theta }{\sin^2 \theta}	\phi 	+
	\lambda \omega \frac{p \cos \theta + q}{\sin^2 \theta} +
	\rho \omega \frac{p + q \cos \theta}{\sin^2 \theta} & = 0 ,
\label{Maxwell_cl_t}\\
	\rho'' -\cot \theta \rho' +
	\cos \theta \lambda^{\prime \prime} -
	\frac{\lambda'}{\sin \theta} +
			\rho \left(\frac{\omega ^2}{4} - q^2\right) +
	\lambda \left(
		\frac{1}{4} \omega ^2 \cos \theta + p q
	\right) -
	\phi\frac{\omega \left(
			p + q \cos \theta
		\right)}{4} & = 0 ,
\label{Maxwell_cl_chi}\\
	\frac{1}{4} \omega  \sin^2 \theta \; \phi ' -
	\lambda' \left(
	p \cos \theta + q
	\right) -
	\rho' \left( p + q \cos \theta
	\right)  & = 0 ,
\label{Maxwell_cl_theta}\\
	\lambda'' -\cot \theta \lambda' +
	\cos \theta \rho^{\prime \prime} -
	\frac{\rho'}{\sin \theta} +
	\lambda \left(\frac{\omega ^2}{4} - p^2\right) +
	\rho \left(
		\frac{1}{4} \omega ^2 \cos \theta + p q
	\right) -
	\phi\frac{\omega \left(
			q + p \cos \theta
		\right)}{4} & = 0 .
\label{Maxwell_cl_varphi}
\end{align}
This set of equations must be regarded as an eigenvalue problem with the eigenfunctions
$\phi_{p q n}, \rho_{p q n}, \lambda_{p q n}$ and the eigenvalue $\omega_{pqn}$,
where the integer $n$ numbers the eigenvalues of $\omega$ for fixed values of the integers $p$ and $q$.

This set of equations has the following discrete symmetries:
\begin{align}
	\phi_{pqn} & = \phi_{qpn}, \quad
	\lambda_{p q n} = \rho_{q p n}, \quad
	\rho_{p q n} = \lambda_{q p n} ,
\label{symmetry_1}\\
	p & \rightarrow - p, \quad
	q \rightarrow - q , \quad
	\omega \rightarrow - \omega .
\label{symmetry_2}
\end{align}
Below we consider regular solutions, as well as singular solutions supported by a pointlike charge and a current located at the points
 $\theta = 0, \pi$.

\subsection{Regular solutions}
\label{CED_reg_sols}

Deriving an analytic solution to the set of equations~\eqref{Maxwell_cl_t}-\eqref{Maxwell_cl_varphi} runs into great difficulty. Therefore, in this subsection, we first find numerical solutions and then show that there are analytic solutions for some particular values of the numbers $p, q$ and of the functions $\phi, \rho$, and $\lambda$.

\subsubsection{Numerical solutions.}

To perform numerical computations, it is necessary to assign the values of the functions and their derivatives at the point $\theta  \ll 1$.
To do this, let us seek a solution in the form
$$
	\phi = \phi_0 \theta^\alpha + \cdots ,
	\lambda = \lambda_0 \theta^\beta + \cdots ,
	\rho_0 = \rho_0 \theta^\beta + \cdots .
$$
Then Eq.~\eqref{Maxwell_cl_theta} yields the following restriction on the parameters $\phi_0, \lambda_0$, and $\rho_0$:
$$
	\phi_0 = \frac{4  (2 + \alpha)}{\omega }
	\left(
		\lambda_0 + \rho_0
	\right)
$$
with
$	\alpha = p + q$ and $\beta = 2+\alpha$.

\begin{figure}[H]
\centering
  \includegraphics[height=4cm]{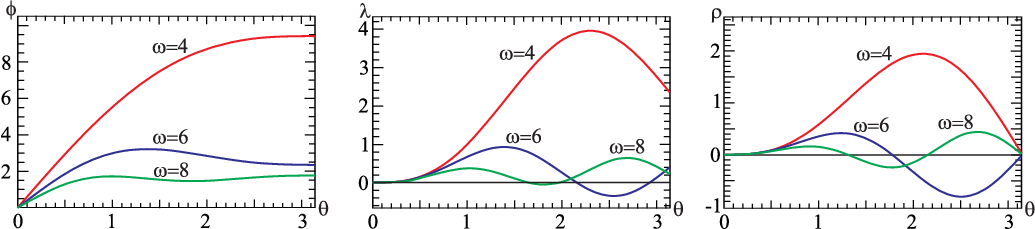}
\caption{The profiles of $A_t = \phi(\theta)$, $A_\varphi = \lambda(\theta)$, and  $A_\chi = \rho(\theta)$
for $\lambda_0 = \rho_0 = 1$.
}
\label{fig_pot}
\end{figure}

\begin{figure}[H]
\centering
  \includegraphics[height=4cm]{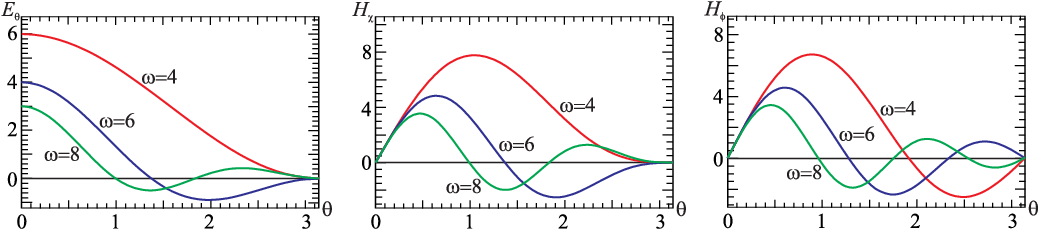}
\caption{The profiles of the electric field $E_\theta = \phi'(\theta)$ and magnetic fields $H_\chi = 2 \frac{\lambda' + \cos \theta \rho'}{\sin \theta}$ and
$H_\varphi = 2 \frac{\rho' + \cos \theta \lambda'}{\sin \theta}$.
For all graphs $\lambda_0 = \rho_0 = 1$.
}
\label{fig_stren}
\end{figure}

Figs.~\ref{fig_pot} and \ref{fig_stren} show the results of numerical calculations of Eqs.~\eqref{Maxwell_cl_t}-\eqref{Maxwell_cl_varphi} for the case of
$p = 0$ and $q = 1$.

The numerical solutions obtained for different $p$ and $q$ permit us to assume that apparently there also exist solutions for all pairs of  $p$ and $q$,
and they are regular for $p + q \geqslant 0$. Thus, we can conclude that the set of equations~\eqref{Maxwell_cl_t}-\eqref{Maxwell_cl_varphi} has apparently a discrete spectrum of solutions
\begin{equation}
	\left( A_\mu \right)_{pqn} = e^{i \left(
		\frac{\omega_{pqn}}{r} t + p \chi + q \varphi
		\right) }
	\left\lbrace
		\phi_{pqn}, r \rho_{pqn}, 0, r \lambda_{pqn}
	\right\rbrace =
	 e^{i \left(
		\frac{\omega_{pqn}}{r} t + p \chi + q \varphi
	\right) } \left( \tilde A_\mu \right)_{pqn} .
\label{Makswell_spectrum}
\end{equation}
According to the property of symmetry~\eqref{symmetry_2}, there also exists the solution
$$
	\left( A_\mu \right)_{pqn} = e^{- i \left(
		\frac{\omega_{pqn}}{r} t + p \chi + q \varphi
		\right) } \left( \tilde A_\mu \right)_{pqn} .
$$

\subsubsection{The case of $p = q = \phi = 0, \rho = \lambda$.}

Here, it is possible to find an analytic solution. In this case the Maxwell equations~\eqref{Maxwell_cl_t}-\eqref{Maxwell_cl_varphi}
reduce to one equation
$$
	\lambda'' - \frac{\lambda'}{\sin \theta} +
	\frac{\omega^2}{4} \lambda = 0,
$$
whose solution is
$$
	\lambda = \frac{c_1}{2} \left( \cos \theta - 1 \right) \,
	_2F_1\left(
		1 - \frac{\omega }{2}, 1+\frac{\omega }{2} ;
		2; \frac{1 - \cos \theta}{2}
	\right) ,
$$
where $_2F_1$ is the hypergeometric function.
This solution is regular for the following eigenvalues of $\omega$:
$$
	\omega_n = \pm 2 n,
$$
where $n$ is integer.

\subsubsection{The case of $p=q$}

Making use of the substitution $(\rho, \lambda) \rightarrow \frac{\omega}{8p} (\tilde \rho, \tilde \lambda)$,
Eqs.~\eqref{Maxwell_cl_t}-\eqref{Maxwell_cl_varphi} can be rewritten as
\begin{align}
	\phi '' + \cot \theta \phi ' -
	\frac{p^2 + q^2 + 2 p q \cos \theta }{\sin^2 \theta}	\phi 	+
	\tilde \lambda \frac{\omega^2}{8}
	\frac{\cos \theta + \frac{q}{p}}{\sin^2 \theta} +
	\tilde \rho \frac{\omega^2}{8}
	\frac{1 + \frac{q}{p} \cos \theta}{\sin^2 \theta} & = 0 , \nonumber
\\
	\tilde \rho'' -\cot \theta \tilde \rho' +
	\cos \theta \tilde \lambda^{\prime \prime} -
	\frac{\tilde \lambda'}{\sin \theta} +
			\tilde \rho \left(\frac{\omega ^2}{4} - q^2\right) +
	\tilde \lambda \left(
		\frac{1}{4} \omega ^2 \cos \theta + p q
	\right) -
	2 \phi p^2 \left(
			1 + \frac{q}{p} \cos \theta
		\right) & = 0 , \nonumber
\\
	\sin^2 \theta \; \phi ' -
	\tilde \lambda'
	\frac{\cos \theta + \frac{q}{p}}{2} -
	\tilde \rho' \frac{1 + \frac{q}{p} \cos \theta}{2}  & = 0 , \nonumber
\\
	\tilde \lambda'' -\cot \theta \tilde \lambda' +
	\cos \theta \tilde \rho^{\prime \prime} -
	\frac{\tilde \rho'}{\sin \theta} +
	\tilde \lambda \left(\frac{\omega ^2}{4} - p^2\right) +
	\tilde \rho \left(
		\frac{1}{4} \omega ^2 \cos \theta + p q
	\right) -
	2 \phi p^2 \left(
			\cos \theta + \frac{q}{p}
		\right) & = 0 .\nonumber
\end{align}

Consider the particular case of $q = p$. In this case
$\tilde \lambda(\theta) = \tilde \rho(\theta)$, and the Maxwell equations take the form
\begin{align}
	 \phi'' + \cot \theta \phi' - p^2
	 \frac{\phi}{\sin^2\left(\frac{\theta }{2}\right)} +
	 \frac{\omega^2}{8} \frac{\tilde \rho}{\sin^2\left(\frac{\theta }{2}\right)}
	 &= 0 ,
\label{maxwell_reduced_10}\\
	\tilde \rho'' - \frac{\tilde \rho'}{\sin \theta} + \frac{\omega ^2}{4} \tilde \rho -
	2 p^2 \phi &= 0 ,
\label{maxwell_reduced_20}\\
	\tilde \rho' + (\cos \theta - 1) \phi' &= 0.
\label{maxwell_reduced_30}
\end{align}
Expressing from Eq.~\eqref{maxwell_reduced_20}
\begin{equation}
	\phi = \frac{\tilde \rho'' - \frac{\tilde \rho'}{\sin \theta} +
	\frac{\omega^2}{4} \tilde \rho}{2 p^2}
\label{phi}
\end{equation}
and substituting it in~\eqref{maxwell_reduced_30}, we have
$$
	\sin^2 \theta \tilde \rho''' - \sin \theta \tilde \rho'' +
	\tilde \rho'
	\left[
		\frac{\omega^2}{4} \sin^2 \theta +
		\left(1 - 2 p^2\right) \cos \theta - 2 p^2
	\right]  = 0.
$$
Its solution for $\tilde \rho'$ gives
\begin{equation}
\begin{split}
	\tilde \rho^\prime =&
 c_1 (1 - \cos \theta )^{p+\frac{1}{2}}
 (1 + \cos \theta)^{-\frac{1}{2}}
 \, _2F_1\left(p - \frac{\omega }{2},
 p+\frac{\omega }{2};
 1 + 2 p; \frac{1-\cos \theta}{2} \right)  	\nonumber
\\
 &
	+c_2 (1 - \cos \theta )^{- p + \frac{1}{2}}
	(1 + \cos \theta)^{-\frac{1}{2}}
 \, {}_2F_1
	 \left(- p- \frac{\omega }{2}, - p + \frac{\omega }{2};
	 1-2 p; \frac{1 - \cos \theta}{2}
 \right) . \nonumber
\end{split}
\end{equation}
Thus, we have two linearly independent solutions, and the second solution can be obtained from the first one on simply replacing $p \rightarrow - p$.
Therefore, it is sufficient to consider only the first solution.

According to Eq.~\eqref{maxwell_reduced_30}, we have the expression 
\begin{equation}
\begin{split}
	\phi^\prime =&
 c_1 (1 - \cos \theta )^{p - \frac{1}{2}}
 (1 + \cos \theta)^{-\frac{1}{2}}
 \, _2F_1\left(p - \frac{\omega }{2},
 p+\frac{\omega }{2};
 1 + 2 p; \frac{1-\cos \theta}{2} \right)  	
\\
 &
	+c_2 (1 - \cos \theta )^{- p - \frac{1}{2}}
	(1 + \cos \theta)^{-\frac{1}{2}}
 \, {}_2F_1
	 \left(- p- \frac{\omega }{2}, - p + \frac{\omega }{2};
	 1-2 p; \frac{1 - \cos \theta}{2}
 \right) .
\label{phi_prime_soln}
\end{split}
\end{equation}
As was already pointed out above, the hypergeometric function  in  \eqref{phi_prime_soln} is regular when
$$
	\omega = \pm 2 (p + l),
$$
where $l$ is an integer. In this case
$$
	{}_2F_1\left(p - \frac{\omega }{2},
	p+\frac{\omega }{2};
	1 + 2 p; \frac{1-\cos \theta}{2} \right) =
	\frac{l!}{
		\left( 1 + 2 p\right)_l
	} P^{\left( 2 p, -1\right) }_l
	\left( \cos \theta\right) .
$$
Thus, for the first independent solution we have the following expressions:
\begin{align}
	\tilde \rho' =&  c_1 \frac{l!}{\left( 1 + 2 p\right)_l }
	(1 - \cos \theta )^{p+\frac{1}{2}}
	(1 + \cos \theta)^{-\frac{1}{2}}
	P^{\left( 2 p, -1\right) }_l (\cos \theta) ,
\label{rho_soln_10}\\
	\phi' =&  c_1 \frac{l!}{\left( 1 + 2 p\right)_l }
	(1 - \cos \theta )^{p - \frac{1}{2}}
	(1 + \cos \theta)^{-\frac{1}{2}}
	P^{\left( 2 p, -1\right) }_l (\cos \theta),
\label{rho_soln_20}\\
	\tilde \rho =&  c_1 \frac{l!}{\left( 1 + 2 p\right)_l }
	\int (1 - \cos \theta )^{p+\frac{1}{2}}
	 (1 + \cos \theta)^{-\frac{1}{2}}
	 P^{\left( 2 p, -1\right) }_l (\cos \theta)
	  d \theta ,\label{rho_soln_30}\\
	\phi = &  c_1 \frac{l!}{\left( 1 + 2 p\right)_l }
	\int (1 - \cos \theta )^{p- \frac{1}{2}}(1 + \cos \theta)^{-\frac{1}{2}}
	 P^{\left( 2 p, -1\right) }_l (\cos \theta) d \theta	.
\label{rho_soln_40}
\end{align}

Taking into account that the Jacobi polynomials can be recast in the form
$$
	P^{(\alpha, \beta)}_n (x) =
	\frac{\Gamma (\alpha + n + 1)}{n! \Gamma (\alpha + \beta + n + 1)}
	\sum \limits_{m=0}^n \binom{n}{k}
	\frac{\Gamma (\alpha + \beta + n + m + 1)}
	{\Gamma (\alpha + m + 1)}
	\left( \frac{x - 1}{2} \right)^m ,
$$
we may integrate the expression in \eqref{rho_soln_30}:
\begin{equation}
\begin{split}
	\tilde \rho_{p p l} = &  -c_1 \frac{l!}{\left( 1 + 2 p\right)_l }
	\int (1 - x)^{p}
	(1 + x)^{-1}
	P^{\left( 2 p, -1\right) }_l (x) d x \nonumber
\\
	= & c_1 \frac{l!}{\left( 1 + 2 p\right)_l }
	\frac{\Gamma (2p + l + 1)}{l! \Gamma (2p + l)}
	\sum \limits_{m=0}^l (-1)^m \binom{l}{k}
	\frac{\Gamma (2p + l + m)}{\Gamma (2p + m + 1)}
	\frac{1}{2^{m+1} (1+m+p)}
	(1-x)^{1+p+m}	
\\
	&
	\times {}_2F_1\left(1,1+m+p;2+m+p;\frac{1-x}{2}\right), \nonumber
\end{split}
\end{equation}
where $p, q=p, l$ are quantum numbers.  Substituting  $\tilde\rho'$ from \eqref{rho_soln_10} in
\eqref{maxwell_reduced_20}, one can find a relation between $\tilde\rho_{p p l}$ and $\phi_{p p l}$,
$$
\frac{\omega^2}{4} \tilde\rho_{p p l} - 2 p^2 \phi_{p p l}	=
	c_1 \frac{l!}{\left( 1 + 2 p\right)_l }
(1-x)^{p+1}\left\{
\frac{\Gamma(2p+l+1)}{2\Gamma(2p+l)}P^{(2p+1, 0)}_{l-1}(x)+
\left[\frac{p}{1-x}+\frac{1}{1+x}-(1+x)
\right]P^{(2p, -1)}_{l}(x)
\right\},
$$
which does not already involve derivatives.

\subsubsection{Another $\mathfrak{Ansatz}$
}

The parent Maxwell equations can be simplified by choosing another
$\mathfrak{Ansatz}$ for the electromagnetic field potentials [which is distinct from that given by Eq.~\eqref{first_ans}]:
\begin{equation}
\begin{split}
	A_\mu = e^{-i  \Omega t} e^{i n \chi} e^{i m \varphi}
	\biggl\{
		\phi, &
		r \left[
			\rho - \lambda \cos \theta +
			\phi \frac{
				\cos \theta \left( p \cos \theta + q \right) + p+q \cos \theta
			}{\omega \sin^2 \theta}
		\right],
		0,\nonumber
\\
	&
	\left.
		r \left[
			\rho - \lambda \cos \theta +
			\phi \frac{
				\cos \theta \left( p \cos \theta + q \right) + p+q \cos \theta
			}{\omega \sin^2 \theta}
		\right]
	\right\rbrace . \nonumber
\end{split}
\end{equation}
This $\mathfrak{Ansatz}$ is chosen so as to maximally simplify the contravariant electromagnetic field tensor $F^{\mu \nu}$.
In this case the Maxwell equations will be
\begin{align}
	\phi'' + \cot \theta \phi' + p \omega  \rho + q \omega  \lambda = & 0 ,
\label{Maxwell_second_10}\\
	\rho'' + \frac{ p+q \cos \theta }	{\omega \sin^2 \theta} \phi''
	-\frac{ p \cos \theta + q}{\omega \sin^3 \theta} \phi'
	+ \frac{\lambda'}{\sin \theta} + 2 \cot \theta \rho'
	- \rho 		
	\frac{
	q^2  + p q \cos \theta -
	\left(\frac{\omega^2}{4} - 1\right) \sin^2 \theta
	}{\sin^2 \theta}  &
\nonumber\\
	+q \lambda \frac{p+q \cos \theta}{\sin^2 \theta}
	= & 0 ,
\label{Maxwell_second_20}\\
	\frac{
		p^2 + q^2 + 2 p q \cos \theta - \frac{\omega^2}{4} \sin^2 \theta
	}{\omega \sin^2 \theta} \phi' +
	p \rho' + q \lambda' +
	\lambda \frac{p+q \cos \theta}{\sin \theta} +
	\rho \frac{p \cos \theta + q}{\sin \theta} = & 0 ,
\label{Maxwell_second_30}\\
	\lambda'' +
	\frac{p \cos \theta + q}{\omega \sin^2 \theta} \phi''
	-\frac{p + q \cos \theta }{\omega \sin^3 \theta} \phi'
	+ 2 \cot \theta \lambda' + \frac{\rho'}{\sin \theta}
	- \lambda
	\frac{
		 p^2 + p q \cos \theta -
		 \left(\frac{\omega^2}{4} - 1\right) \sin^2 \theta
	}{\sin^2 \theta}  &
\nonumber\\
	+ p \rho \frac{p \cos \theta + q}{\sin^2 \theta}
	= & 0 .
\label{Maxwell_second_40}
\end{align}
One can find  $\phi''$  from Eq.~\eqref{Maxwell_second_10} and substitute it into Eqs.~\eqref{Maxwell_second_20} and \eqref{Maxwell_second_40} to yield
\begin{align}
	\rho'' -
	\frac{4 p \cos \theta +
	q \left( \cos 2 \theta + 3 \right) }{2 \omega \sin^3 \theta} \phi'
	+ \frac{\lambda'}{\sin \theta} + 2 \cot \theta \rho' -
	\rho
	\frac{
		p^2 + 2 p q \cos \theta + q^2 -
		\left(\frac{\omega^2}{4} - 1\right) \sin^2 \theta
	}{\sin^2 \theta} = & 0 ,
\label{lambda_rho_eqn_10}\\
	\lambda'' -
	\frac{4 q \cos \theta +
	p \left( \cos 2 \theta + 3 \right) }{2 \omega \sin^3 \theta} \phi'
	+ 2 \cot \theta \lambda' + \frac{\rho'}{\sin \theta} -
	\lambda
	\frac{
		p^2 + 2 p q \cos \theta + q^2 -
		\left(\frac{\omega^2}{4} - 1\right) \sin^2 \theta
	}{\sin^2 \theta} = & 0 ,
\label{lambda_rho_eqn_20}
\end{align}
where $\phi'$ is determined from Eq.~\eqref{Maxwell_second_30} as
$$
	\phi' =
	- \omega  \sin \theta \frac{
		\sin \theta \left(p \rho' + q \lambda' \right) +
		\lambda (p + q \cos \theta) +
		\rho (p \cos \theta + q)
	}{
	\left(p^2 + q^2\right) + 2 p q \cos \theta -
	\frac{\omega ^2}{4}\sin^2 \theta
	} .
$$

The equations~\eqref{lambda_rho_eqn_10} and \eqref{lambda_rho_eqn_20}  can be approximately solved for the case of large $p$ and $q$
assuming that the eigenvalue $\omega$ has the same order as that of the quantum numbers $p$ and $q$:
\begin{equation}
	\omega \approx (p, q), \quad
	\left( p, q, \omega \right) \gg 1 .
\label{conditions}
\end{equation}
We also assume that
$$
	\rho' \approx (p, q, \omega) \rho, \quad
	\lambda' \approx (p, q, \omega) \lambda, \quad
	\phi' \approx \left( \rho', \lambda' \right),
	\text{ and }
	\left( \rho'', \lambda''\right) \approx
	\left( p^2, q^2, \omega^2 \right) \left( \rho, \lambda\right) .
$$
In this case it is possible to neglect the second, third, and fourth terms in the left-hand sides of Eqs.~\eqref{lambda_rho_eqn_10} and \eqref{lambda_rho_eqn_20}.
As a result, we have the same equation for $\lambda$ and $\rho$,
$$
	\rho'' -
	\rho
	\frac{
		p^2 + 2 p q \cos \theta + q^2 -
		\frac{\omega^2}{4}  \sin^2 \theta
	}{\sin^2 \theta} = 0 .
$$
This equation can be approximately solved, subject to the conditions \eqref{conditions}, to yield
\begin{equation}
\begin{split}
	\left( \rho, \lambda \right)  \approx & c_1
	\left( 1 - \cos \theta\right)^{\frac{p + q}{2}}
	\left( 1 + \cos \theta\right)^{\frac{p - q}{2}}
	{}_2F_1 \left(
		p - \frac{\omega}{2}, p + \frac{\omega}{2};
		p + q ; \frac{1 - \cos \theta}{2}
	\right)\nonumber
\\
	+ & c_2
	\left( 1 - \cos \theta\right)^{-\frac{p + q}{2}}
	\left( 1 + \cos \theta\right)^{\frac{p - q}{2}}
	{}_2F_1 \left(
		- q - \frac{\omega}{2}, - q + \frac{\omega}{2};
		- p - q ; \frac{1 - \cos \theta}{2}
	\right) . \nonumber
\end{split}
\end{equation}
Here, we have two linearly independent solutions, and the second solution can be obtained from the first one using the replacements
$p \rightarrow -q, q \rightarrow -p$; for this reason, we consider further only the first independent solution.

As in the cases considered above, the hypergeometric function will be regular at the points $\theta = 0, \pi$ if
$$
	\omega = \pm 2 \left( p + l \right).
$$
In this case the hypergeometric function is the Jacobi polynomial defined according to \eqref{g_sol_D_120}:
$$
	\left( \rho_{p q l}, \lambda_{p q l} \right)  \approx \tilde c_1
	\left( 1 - \cos \theta\right)^{\frac{p + q}{2}}
	\left( 1 + \cos \theta\right)^{\frac{p - q}{2}}
	P^{(p+q-1, p-q+l)}_l \left( \cos \theta \right) .
$$

\subsubsection{Singular solutions
}

Singular solutions are supported by a pointlike charge and a pointlike current,  and for them $p = q = \omega = 0$.
In this case the Maxwell equations~\eqref{Maxwell_cl_t}-\eqref{Maxwell_cl_varphi} take the form
\begin{align}
	\frac{1}{\sin \theta} \left(
		\sin \theta \phi^\prime
	\right)^\prime & = 0 ,
\nonumber\\
	\left(
		\frac{\rho^\prime}{\sin \theta}
	\right)^\prime +
	\left(
		\cot \theta \lambda^\prime
	\right)^\prime & = 0 ,
\nonumber\\
	\left(
		\frac{\lambda^\prime}{\sin \theta}
	\right)^\prime +
	\left(
		\cot \theta \rho^\prime
	\right)^\prime & = 0
\nonumber
\end{align}
with the solutions
\begin{align}
	\phi & = a_1 - a_2 \arctanh (\cos \theta) =
	a_1 + a_2 \ln \sin \left( \frac{\theta}{2}\right) -
	a_2 \ln \cos \left( \frac{\theta}{2}\right),
\label{sol_10}\\
	\lambda &= c_3 - \left( c_1 - c_2 \right) \ln \left[
			\sin \left(\frac{\theta}{2}\right)
		\right] - \left( c_1 + c_2\right)
		\ln \left[
			\cos \left(\frac{\theta}{2}	\right)
		\right]  ,
\label{sol_20}\\
	\rho &= c_4 + \left( c_1 - c_2 \right) \ln \left[
		\sin \left(\frac{\theta}{2}\right)
	\right] - \left( c_1 + c_2\right)
	\ln \left[
		\cos \left(\frac{\theta}{2}	\right)
	\right]  ,
\label{sol_30}
\end{align}
where $a_i$ and $c_j$ are integration constants.
In the above expressions, the terms with
$
	\ln \left[ \sin \left(\frac{\theta}{2}\right) \right]
$
describe the electric and magnetic fields created by the pointlike charge and current located at the point $\theta = 0$
where the potentials diverge. Similarly, one can regard the terms with
$
	\ln \left[ \cos \left(\frac{\theta}{2}\right) \right]
$ as describing the electric and magnetic fields created by the pointlike charge and current located at the point $\theta = \pi$.

The components of the electric, $E_\theta = - \partial_\theta \phi$, and magnetic, $H_{\chi, \varphi}$, fields are
\begin{align}
	E_\theta &= - \frac{q}{2} \tan \left( \frac{\theta}{2}\right) -
	\frac{q}{2} \cot \left( \frac{\theta}{2}\right) ,
\label{el_field}\\
	H_\chi &= 2 \frac{\lambda' + \cos \theta \rho'}{\sin \theta}=2 c_2 ,
\label{mg_field_chi}\\
	H_\varphi &= 2 \frac{\rho' + \cos \theta \lambda'}{\sin \theta} =2 c_1 .
\label{mg_field_phi}
\end{align}
The first term in the right-hand side of Eq.~\eqref{el_field} describes the electric field created by the pointlike charge $q$ located at the point $\theta = \pi$, and
the second term~--  the field created by the pointlike charge $q$ located at the point $\theta = 0$. Thus the solutions~\eqref{sol_10}-\eqref{sol_30} and \eqref{el_field}-\eqref{mg_field_phi}
describe the scalar/vector potentials and electric/magnetic fields created by two pointlike charges and currents located at the points $\theta = 0, \pi$.

Summarizing the results obtained in this section, we have studied some particular cases for Eqs.~\eqref{Maxwell_cl_t}-\eqref{Maxwell_cl_varphi} and shown that in these cases
regular classical solutions to the Maxwell equations on the Hopf bundle are determined by three quantum numbers.
This enables us to suppose that a general solution to the Maxwell equations on a three-dimensional sphere is also determined by three quantum numbers.
The fact that the spectrum of the solutions on a three-dimensional sphere is discrete and is numbered by three integers has the natural explanation that the sphere is a compact
space (this gives rise to the discreteness of the spectrum), and the dimension of the sphere equal to 3 gives rise to three numbers related to the spectrum of the solutions.

\section{Quantization of linear fields
}
\label{quantum_lin_fields}

In this section we consider the quantization of the Dirac field and Maxwell's electrodynamics  in $\mathbb{R} \times S^3$ spacetime where a spatial cross-section $S^3$
is the Hopf bundle $S^3 \rightarrow S^2$.

The distinctive feature of the field theories under consideration on the Hopf bundle is that a three-dimensional sphere  $S^3$ is a compact space.
This results in the fact that the noninteracting field systems in question (spinor Dirac field and Maxwell's electrodynamics) have discrete spectra of solutions;
in both cases, this enables us to write a general solution as a discrete sum over the corresponding eigenvalues. This is the principle difference compared to Minkowski space
where a general solution is given by the Fourier integral. The replacement of the integral by the sum should lead to a considerable simplification of the quantization process.

\subsection{Quantization of the Dirac field
}
\label{quant_free_Dirac}

Consistent with the discrete spectrum of the solutions~\eqref{g_sol_D_210},
a general solution to the Dirac equation in the classical case can be represented as a sum of these solutions numbered by the integers $m,n$, and $l$.
Then the field operators $\hat \psi$ and $\hat \psi^\dagger$ can be written in the form
\begin{align}
	\hat \psi &= \sum_{n}
	\sum_{\left| m \right| \leqslant n} \sum_{l}
	\left[
		\hat b_{n m l} e^{
			i \left( - \Omega_{n l} t + n \chi + m \varphi\right)
		}
		\Xi_{nml} \left( \cos \theta\right) +
		\left( \hat c_{n m l}\right)^\dagger e^{
			i \left( - \Omega_{-n, l} t - n \chi - m \varphi\right)
		}
		\tilde \Xi_{nml} \left( \cos \theta\right)
	\right] , \nonumber
\\
	\hat \psi^\dagger &= \sum_{n}
	\sum_{\left| m \right| \leqslant n}  \sum_{l}
	\left[
		\left( \hat b_{n m l}\right)^\dagger e^{
			- i \left( - \Omega_{n l} t + n \chi + m \varphi\right)
		}
		\Xi_{nml} \left( \cos \theta\right) +
		\hat c_{n m l} e^{
			-i \left( - \Omega_{-n, l} t - n \chi - m \varphi\right)
		}
		\tilde \Xi_{nml} \left( \cos \theta\right)
	\right], \nonumber
\end{align}
where the spinors $\Xi_{nml},\tilde \Xi_{nml}$ are
$$
	\Xi_{nml}(\cos \theta)=
		\begin{pmatrix}
		\Theta_{1, nml}  \\
		\Theta_{2, nml}
	\end{pmatrix}, \quad
	\tilde \Xi_{nml}(\cos \theta)=
\begin{pmatrix}
	\Theta_{2, -n, -m, l}  \\
	\Theta_{1, -n, -m, l}
\end{pmatrix}
$$
and defined according to Eqs.~\eqref{g_sol_D_190} and \eqref{g_sol_D_200}. The energy given by Eq.~\eqref{g_sol_D_110} is
$$
 \tilde \Omega_{n l} = \pm (n + l)  , \text{ where }
 l = 0,  1, \ldots
$$
and
\begin{equation*}
	\left| n \right| \geqslant 1 ,
	\left| m \right| \leqslant \left| n \right| .
\end{equation*}
The operator $\hat b_{n m l}$ describes the annihilation of a particle with the energy $r \Omega_{n l} = \pm(n + l)$, and the quantum numbers $n,m, l$. Correspondingly, the operator
$\left( \hat b_{n m l}\right)^\dagger$ describes the creation of such a particle. Similarly, the operator $\hat c_{n m l}$ describes the annihilation of a particle with the energy
$r \Omega_{-n, l} = \pm(- n + l)$, and the quantum numbers $n,m, l$, and the operator
$\left( \hat c_{n m l}\right)^\dagger$ describes the creation of such a particle.

We impose the following anticommutation relations on these operators:
\begin{align}
	\left\lbrace
		\hat b_{n m l} , \left( \hat b_{p q r}\right) ^\dagger
		\right\rbrace & = f_{nml} \delta_{mp} \delta_{nq}
	\delta_{lr},
\label{quantum_Dirac_60}\\
	\left\lbrace
		\hat c_{n m l} , \left( \hat c_{p q r}\right) ^\dagger
		\right\rbrace & = f_{nml} \delta_{mp} \delta_{nq}
	\delta_{lr},
\label{quantum_Dirac_70}\\
	\left\lbrace
		\hat b_{n m l} , \hat b_{p q r}
		\right\rbrace & =
	\left\lbrace
		\hat c_{n m l} , \hat c_{p q r}
	\right\rbrace =
	\left\lbrace
		\left( \hat b_{n m l}\right) ^\dagger  ,
		\left( \hat b_{p q r}\right) ^\dagger
	\right\rbrace =
	\left\lbrace
		\left( \hat c_{n m l}\right) ^\dagger  ,
		\left( \hat c_{p q r}\right) ^\dagger
	\right\rbrace = \dots = 0,
\label{quantum_Dirac_80}
\end{align}
where $f_{nml}$ is a numerical factor possibly depending on $n$, $m$, and $l$, and this factor is chosen such that the infinite sum over $n$ and $l$
in the propagator \eqref{quantum_Dirac_90} would be convergent.
The fermion propagator is defined as usual by the expression
\begin{equation}
\begin{split}
	i S_{\alpha \beta} & \left(
		t - t', \chi - \chi', \theta,\theta',	\varphi - \varphi'
	\right)
	=
	\left\lbrace
		\hat \psi_\alpha \left( t, \chi, \theta, \varphi \right) ,
		\hat{\bar \psi}_\beta
		\left( t^\prime, \chi^\prime, \theta^\prime, \varphi^\prime \right)
	\right\rbrace
\\
	&
	= \sum_{n} \sum_{p}
	\sum_{\left| m \right| \leqslant n} \sum_{\left| q \right| \leqslant p}
	\sum_l \sum_r
\\
	&
	\biggl[
		\left\lbrace
			\hat b_{n m l} , \left( \hat b_{p q r}\right)^\dagger
		\right\rbrace
		e^{
			i \left(
				- \Omega_{nl} t + \Omega_{pr} t' + n \chi - p \chi'
				+ m \varphi - q \varphi'
			\right)
 	}
 	\left[ \Xi_{nml}\left( \cos \theta\right) \right]_\alpha
 	\left[ \tilde \Xi_{pqr} \left( \cos \theta'\right)\right]_\beta
\\
	&
	+
		\left\lbrace
			\left( \hat c_{nml}\right)^\dagger , \hat c_{pqr}
		\right\rbrace
		e^{
			i \left(
				\Omega_{nl} t - \Omega_{pr} t' - n \chi + p \chi'
				- m \varphi + q \varphi'
			\right)
 	}
 	\left[ \Xi_{nml} \left( \cos \theta\right) \right]_\alpha
 	\left[ \tilde \Xi_{pqt}\left( \cos \theta'\right) \right]_\beta
	\biggl]
\\
	&
	= \sum_{n}
	\sum_{\left| m \right| \leqslant n} \sum_{l} f_{nml}
	\left\lbrace
		e^{
			i \left[
				- \Omega_{nl} (t - t') + n (\chi - \chi')
				+ m (\varphi - \varphi')
			\right]
		}
 	\left[ \Xi_{nml} \left( \cos \theta\right)\right]_\alpha
 	\left[ \tilde \Xi_{nml} \left( \cos \theta'\right)\right]_\beta
	\right.
\\
	&
	\left.
+	e^{
		- i \left[
			- \Omega_{nl} (t - t') + n (\chi - \chi')
			+ m (\varphi - \varphi')
		\right]
 	}
 	\left[ \Xi_{nml}\left( \cos \theta\right) \right]_\alpha
 	\left( \tilde \Xi_{nml}\left( \cos \theta'\right) \right]_\beta
	\right\rbrace
\\
	& = 2 \sum_{n}
	\sum_{\left| m \right| \leqslant n} \sum_l f_{nml}
	\left\lbrace
	\cos
		\left[
			- \Omega_{nl} (t - t') + n (\chi - \chi')
			+ m (\varphi - \varphi')
		\right]
 	\tilde S_{\alpha \beta; nml}\left(
 		\cos \theta, \cos \theta'
 	\right)
 	\right\rbrace ,
\label{quantum_Dirac_90}
\end{split}
\end{equation}
where
$
	\tilde S_{\alpha \beta;nml} =
 	\left[ \Xi_{nml}\left( \cos \theta\right) \right]_\alpha
	\left( \tilde \Xi_{nml}\left( \cos \theta'\right) \right]_\beta	
$
is a matrix. Note that this propagator is not translationally invariant, since it contains the product
$
F^{\left(s, \alpha, \beta \right) }_{nml} \left( \cos \theta\right)
 	F^{\left(s, \alpha, \beta \right) }_{nml} \left( \cos \theta'\right)
$.

To calculate the Hamilton operator, let us write out its density
$$
	\mathcal{H} = \bar \psi \left(
		- i \gamma^k \nabla_k \psi
	\right) .
$$
Then, after standard calculations, we arrive at the following expression:
$$
	H = \sum_s \sum_{n}	\sum_{\left| m \right| \leqslant n} \sum_l
	\Omega_{nl}
	\left[
		\left( \hat b_{nml}^s\right)^\dagger \hat b_{nml}^s -
		\left( \hat c_{nml}^s\right)^\dagger \hat c_{nml}^s - 1
	\right] .
$$
The question of whether the last term in the square brackets leads to the divergence or not requires a special study, since $\Omega_n$
can be both positive and negative.

\subsection{Quantization of Maxwell's electrodynamics
}

In Sec.~\ref{CED_reg_sols}, the numerical study of solutions within classical electrodynamics has been carried out, and we gave arguments to claim that
these solutions
form a discrete spectrum. The same situation takes place for the Dirac equation as well. The reason for that is that a three-dimensional sphere on the
Hopf bundle $S^3 \rightarrow S^2$ is a compact space; as a result, the spectra of solutions of the Dirac and Maxwell equations are discrete.

For such a case, the operator of the electromagnetic field four-potential can be written in the form
$$
	\hat A_\mu = \sum_{p, q, n}
	\left[
		\hat a_{pqn} e^{i \left(
			\frac{\omega_{pqn}}{r} t + p \chi + q \varphi
		\right) } \left( \tilde A_\mu \right)_{pqn}(\theta) +
		\left( \hat a_{pqn} \right)^\dagger
		e^{ - i \left(
		\frac{\omega_{pqn}}{r} t + p \chi + q \varphi
	\right) } \left( \tilde A_\mu \right)_{pqn}(\theta)
	\right] =
	\hat A_\mu^{(+)} + \hat A_\mu^{(-)}
$$
with the following standard commutation relations for the creation,
 $\left( \hat a_{pqn} \right)^\dagger$, and annihilation,  $\hat a_{pqn}$, operators of the quantum state $\left| pqn \right\rangle$:
$$
	\left[
		\hat a_{pqn} , \left( \hat a_{rsm} \right)^\dagger
	\right] = f_{p q n}
	\delta_{pr} \delta_{qs} \delta_{nm} \quad
	\text{(all other commutators are zero)}.
$$
As for the anticommutation relations~\eqref{quantum_Dirac_60} and \eqref{quantum_Dirac_70}, we have introduced here the numerical factor $f_{pqn}$, which will possibly be needed to ensure the finiteness of the sum over the quantum states $p q n$. The momentum operators conjugate to the potential  $A_\mu$
are defined as
\begin{align}
	\hat \pi^0 & = 0 ,
\nonumber\\
	\hat \pi^\chi = - \hat F^{0 \chi} = \hat E^\chi = &
	\frac{- 4 i}{r^2}\sum_{p, q, n}
	\left[ \hat a_{pqn}
	e^{i \left(
		\frac{\omega_{pqn}}{r} t + p \chi + q \varphi
	\right)} +
	\left( \hat a_{pqn} \right)^\dagger
	e^{- i \left(
	\frac{\omega_{pqn}}{r} t + p \chi + q \varphi
	\right)}	
	\right]\nonumber
\\
	\times &
	\frac{
		\omega \left( \lambda_{pqn} \cos \theta + \rho_{pqn} \right)  -
		\phi_{pqn} \left( p + q \cos \theta \right)
	}{\sin^2 \theta},
\nonumber\\
	\hat \pi^\theta = - \hat F^{0 \theta} = \hat E^\theta = &
	\frac{4}{r^2} \sum_{p, q, n}
	\left[ \hat a_{pqn}
		e^{i \left(
			\frac{\omega_{pqn}}{r} t + p \chi + q \varphi
	\right)} +
	\left( \hat a_{pqn} \right)^\dagger
		e^{- i \left(
			\frac{\omega_{pqn}}{r} t + p \chi + q \varphi
	\right)}	
	\right] \phi^\prime ,
\nonumber\\
	\hat \pi^\varphi = - \hat F^{0 \varphi} = \hat E^\varphi = &
	\frac{- 4 i}{r^2}\sum_{p, q, n}
	\left[ \hat a_{pqn}
	e^{i \left(
		\frac{\omega_{pqn}}{r} t + p \chi + q \varphi
	\right)} +
	\left( \hat a_{pqn} \right)^\dagger
	e^{- i \left(
	\frac{\omega_{pqn}}{r} t + p \chi + q \varphi
	\right)}	
	\right]\nonumber
\\
	\times &
	\frac{
		\omega \left( \lambda_{pqn} + \rho_{pqn} \cos \theta \right)  -
		\phi_{pqn} \left( p + q \cos \theta \right)
	}{\sin^2 \theta} .
\nonumber
\end{align}

Let us now calculate the commutators:
\begin{align}
	\left[
		\hat A_\mu^{(+)} \left( t, \chi, \theta, \varphi \right) ,
		\hat A_\nu^{(-)} \left( t', \chi', \theta', \varphi' \right)
	\right] = & \sum_{p, q, n}
	e^{i \left[
		\frac{\omega_{pqn}}{r} \left( t - t' \right) +
		p \left( \chi - \chi'\right) +
		q \left( \varphi - \varphi'\right)
	\right]} \left( \tilde A_\mu \right)_{pqn}(\theta)
	\left( \tilde A_\nu \right)_{pqn}(\theta')
\nonumber\\
	=&
	- i G^{(+)}_{\mu \nu} \left(
		t - t', \chi - \chi', \theta, \theta', \varphi - \varphi'
	\right) ,
\nonumber\\
	\left[
		\hat A_\mu^{(-)} \left( t, \chi, \theta, \varphi \right) ,
		\hat A_\nu^{(+)} \left( t', \chi', \theta', \varphi' \right)
	\right] = &
	-	\left[
		\hat A_\mu^{(+)} \left( t, \chi, \theta, \varphi \right) ,
		\hat A_\nu^{(-)} \left( t', \chi', \theta', \varphi' \right)
	\right]
\nonumber\\
	=&
	- i G^{(-)}_{\mu \nu} \left(
		t - t', \chi - \chi', \theta, \theta', \varphi - \varphi'
	\right) .
\nonumber
\end{align}
It must be mentioned here that the Green functions $G^{(\pm)}_{\mu \nu}$ are not translationally invariant, since they contain the product
$
	\left( \tilde A_\mu \right)_{pqn}(\theta)
	\left( \tilde A_\nu \right)_{pqn}(\theta')
$. These expressions permit us to calculate the Feynman Green's function
\begin{equation}
	G_{F, \mu \nu} = - \Theta\left( t - t'\right)
	G^{(+)}_{\mu \nu} \left(
		t - t', \chi - \chi', \theta, \theta', \varphi - \varphi'
	\right) +
	\Theta\left( t' - t \right)
	G^{(-)}_{\mu \nu} \left(
		t - t', \chi - \chi', \theta, \theta', \varphi - \varphi'
	\right)	.
\label{3_B_90}
\end{equation}

The results obtained in this section concerning the quantization of free Dirac and Maxwell fields need to be compared with the usual quantization in a box
in Minkowski spacetime. The main difference is that in the box eigenfunctions are plain waves (because  the spacetime is locally flat).
On a sphere, plain waves cannot be eigenfunctions of the Dirac equation, since the spacetime has a nonzero curvature.
For this reason, the propagators are not translationally invariant; they are functions of  $\theta, \theta'$ but not functions of their difference, $(\theta - \theta')$.

\section{Nonperturbative quantization of Maxwell's electrodynamics coupled to a spinor field
}
\label{NP_QED}

In this section we suggest a method of nonperturbative quantization of Maxwell-Dirac theory on the Hopf bundle.
As we saw in Sec.~\ref{quantum_lin_fields}, a fact of compactness of a three-dimensional sphere $S^3$ very much simplifies the procedure of quantization of free fields:
when quantizing, no Dirac delta functions appear, since they are replaced by the Kronecker symbols. This would lead us to expect that the
quantization of interacting fields on a sphere $S^3$ will be considerably simplified as well. Notice also that there appears a considerable difference in
the behavior of quantum fields on a sphere $S^3$ and in Minkowski space: the propagators of free fields on the compact space are not translationally invariant.

According to Heisenberg~\cite{heis}, the process of nonperturbative quantization consists in that a set of equations describing interacting fields [in our case, these are Eqs.~\eqref{D_10} and \eqref{D_15}]
is written in the operator form
\begin{align}
	i \gamma^\mu \hat \psi_{;\mu} =& 0 ,
\label{NP_10}\\
	\frac{1}{\sqrt{-g}}
	\frac{\partial }{\partial x^\nu}
	\left( \sqrt{-g} \hat F^{\mu \nu} \right) =& - 4 \pi \hat j^\mu,
\label{NP_20}
\end{align}
where the covariant derivative of the spinor field, the operators of the field strength and of the current
 are defined as
$	\hat \psi_{;\mu} =  \left[\partial_{ \mu} +
	1/8\, \omega_{a b \mu}\left(
		\gamma^a  \gamma^b -
		\gamma^b  \gamma^a\right) -
		i e \hat A_\mu
	\right] \hat \psi ,
	\hat F_{\mu \nu} =  \partial_\mu \hat A_\nu -
	\partial_\nu \hat A_\mu$, and $\hat j^\mu =  e \hat{\bar \psi} \gamma^\mu \hat \psi$, respectively.

It is worth to point out that the above equations involve the operators of the \emph{interacting} fields $\hat \psi$ and $\hat A_\mu$,
whose properties differ from those of free, noninteracting fields considered in the previous section.
Notice also the presence of the nonlinear quantities $\hat A_\mu \hat \psi$ and $\hat{\bar \psi} \gamma^\mu \hat \psi$.
The former describes the interaction between the fields, and the latter is the source of the electromagnetic field. These nonlinear
quantities do not allow to quantize the fields $\psi$ and  $A_\mu$, as was done in Sec.~\ref{quantum_lin_fields}.
For free fields, we have discrete spectra of solutions, a linear combination of which permits one to find any solution to the Maxwell or Dirac equations.
With such spectra in hand, one can quantize the fields by writing the operators
$\hat \psi$ and $\hat A_\mu$ as a superposition of solutions of the discrete spectrum and by introducing the creation and annihilation operators of the corresponding quantum states which are the coefficients before each such solution. In Minkowski space, such operators are called the creation/annihilation operators for particles. But in our case we cannot speak of particles, since the functions \eqref{g_sol_D_210} and \eqref{Makswell_spectrum} do not correspond to plane waves.

When quantizing free fields, we saw that the propagators \eqref{quantum_Dirac_90} and \eqref{3_B_90} are ordinary (not distribution) functions.
These propagators do not involve Dirac delta functions, and therefore the product of two operators at one point is well defined. This was demonstrated for the propagator
$S_{\alpha \beta}(0,0,\theta,\theta,0)$ in Sec.~\ref{quant_free_Dirac}.
It is reasonable to expect that this property also persists for the operators of interacting fields. Hence, the products of the
\emph{interacting} operators $\hat A_\mu \hat \psi$ and $\hat{\bar \psi} \gamma^\mu \hat \psi$
for the discrete spectrum on the Hopf bundle are well defined, in contrast to the same product of operators of
\emph{free} fields in the case of perturbative quantization in Minkowski space.

According to Ref.~\cite{heis}, the main idea of nonperturbative quantization consists in that the operator equations \eqref{NP_10} and \eqref{NP_20} are replaced by an infinite
system for all Green's functions. The first equations are quantum average of Eqs.~\eqref{NP_10} and \eqref{NP_20}. They contain the Green functions
$
	G(A_\mu(x), \psi(y)) = \left\langle \hat A_\mu (x)
	\hat \psi (y)
	\right\rangle
$ and
$
	G(\bar \psi(x), \psi(y)) = \left\langle
	\hat{\bar \psi}(x) \hat \psi(y)
	\right\rangle
$. In order to close the set of equations, it is necessary to derive equations for these Green functions. To do this, the operator equations \eqref{NP_10} and \eqref{NP_20}
are multiplied by the corresponding operators and are averaged; this is done an infinite number of times. As a result, one arrives at the following infinite set of equations:
\begin{align}
	i \gamma^\mu \left\langle \hat \psi_{;\mu} \right\rangle =& 0 ,
\label{NP_60}\\
	\frac{1}{\sqrt{-g}}
	\frac{\partial }{\partial x^\nu}
	\left( \sqrt{-g}
		\left\langle \hat F^{\mu \nu} \right\rangle
	\right) =& - 4 \pi \left\langle \hat j^\mu \right\rangle ,
\label{NP_70}\\
	i \gamma^\mu
	\left\langle \hat{\bar \psi} \hat \psi_{;\mu}
	\right\rangle =& 0 ,
\label{NP_80}\\
	i \gamma^\mu
	\left\langle \hat A_\nu \hat \psi_{;\mu}
	\right\rangle =& 0 ,
\label{NP_90}\\
	\frac{1}{\sqrt{-g}}
	\frac{\partial }{\partial x^\nu}
	\left( \sqrt{-g}
		\left\langle \hat F^{\mu \nu} \hat \psi \right\rangle
	\right) =& - 4 \pi \left\langle \hat j^\mu \hat \psi \right\rangle ,
\label{NP_100}\\
	\ldots
\label{NP_110}
\end{align}
Note that the Green function $G(A_\mu(x), \psi(y))$ is a function of two variables, and for its definition we need two equations~\eqref{NP_90} and \eqref{NP_100}.
After adding the equations \eqref{NP_80}-\eqref{NP_100}, there appear new Green's functions
$
	\left\langle
		\hat{\bar \psi} \hat A_\mu \hat \psi
	\right\rangle
$,
$
	\left\langle\hat A_\nu \hat A_\mu \hat \psi\right\rangle
$, and
$
	\left\langle\hat{\bar \psi} \hat \psi \hat \psi\right\rangle
$, for which one has to write out new equations, and so on  an infinite number of times  [this is denoted by Eq.~\eqref{NP_110}].

It is evident that this infinite set of equations cannot be solved explicitly and analytically. Therefore, the question arises as to whether it is possible to find its approximate solution.
The problem of how to cut off an infinite set of equations is called \emph{the closure problem}, and it is well known in turbulence modeling (see, e.g., the textbook~\cite{Wilcox}).
In that case, the Navier-Stokes equation is averaged, and it is known as the Reynolds-averaged Navier-Stokes equation. But this equation contains an unknown
quantity~-- the Reynolds-stress tensor, for which one has to have an extra equation called the Reynolds-stress equation, which in turn contains more unknown functions, and so on.

For better understanding of the situation, it is useful to consider some simple example illustrating this process. Consider the case where, to solve the set of equations \eqref{NP_60}-\eqref{NP_110}, 
one can employ the simplified  $\mathfrak{Ansatze}$ \eqref{D_30} for the spinor field and \eqref{D_35} for the potential of the electromagnetic field:
\begin{equation}
		\Theta_{2} = \pm \Theta_{1} = \pm \Theta, \quad
		\tilde \Omega_{n 0} = \frac{1}{4} - \frac{r \Omega}{2} = \pm n,
		\quad
		\phi = \mp 2 \rho .
\label{NP_115}
\end{equation}
This case is a special case of the general $\mathfrak{Ansatze}$ \eqref{D_30} and it corresponds to the particular solution for
$l = 0$ found in Ref.~\cite{Dzhunushaliev:2020dom}.

Replacing the functions by operators, we then have
$$
	\hat \psi_{nm 0 } = e^{-i  \Omega t} e^{i n \chi} e^{i m \varphi}
	\hat \Theta_{nm0 }(\theta)
	\begin{pmatrix}
		1  \\
		\pm 1  \\
		0  \\
		0
	\end{pmatrix} ,
\quad
	\hat A_\mu = \left\lbrace
		\hat \phi(\theta), r \hat \rho(\theta) ,0, r \hat \lambda(\theta)
	\right\rbrace  .
$$
(In what follows we drop the indices $n,m,0$ by $\hat \Theta$ for brevity.)
In this case
we obtain the following equations coming from the quantum equations~\eqref{NP_60}-\eqref{NP_110}:
\begin{align}
	\hat \Theta^\prime \mp \hat \Theta
	\left[
		\cot \theta \left(\mp \frac{1}{2} + n + er \hat \rho\right) +
		\frac{1}{\sin \theta}\left(m + er \hat \lambda\right)
	\right] &= 0 ,
\label{NP_140}\\
	\frac{1}{\sin \theta} \left(
			\sin \theta \hat \phi^\prime
		\right)^\prime &=
		- e r^{3/2} \frac{\hat \Theta^2}{2}  ,
\label{NP_150}\\
	\frac{1}{\sin \theta} \left(
		\sin \theta {\hat\rho}^\prime
	\right)^\prime &= \pm e r^{3/2} \frac{\hat \Theta^2}{4}  ,
\label{NP_160}\\
	\hat \lambda^\prime &= -  \hat \rho^\prime \cos \theta .
\label{NP_170}
\end{align}
After quantum averaging, the first equation~\eqref{NP_140} will be the equation for $\left\langle \hat \Theta \right\rangle$,
the second one~-- for $\left\langle \hat \phi \right\rangle$, the third one~-- for
$\left\langle \hat \rho \right\rangle$, and the fourth one~-- for
$\left\langle \hat \lambda \right\rangle$. But these equations contain the following new Green's functions:
$$
	G_{\Theta \rho}(\theta_1, \theta_2) =
	\left\langle \hat \Theta(\theta_1) \hat \rho(\theta_2) \right\rangle, \quad
	G_{\Theta \lambda}(\theta_1, \theta_2) =
	\left\langle \hat \Theta(\theta_1) \hat \lambda(\theta_2) \right\rangle, \quad
	G_{\Theta \Theta}(\theta_1, \theta_2) =
	\left\langle \hat \Theta(\theta_1) \hat \Theta(\theta_2) \right\rangle ,
$$
for which one must have their own equations. The equation for the Green function $G_{\Theta \rho}(x, y)$
can be obtained by multiplying the operator equation \eqref{NP_140} on the right by $\hat \rho$ and by performing the quantum averaging,
$$
	G_{\bar \Theta \rho}^\prime \mp G_{\Theta \rho}
	\left[
		\cot \theta \left(\mp \frac{1}{2} + n \right) +
		\frac{m}{\sin \theta}
	\right]
	\mp er \left\langle
		\hat \Theta \hat \rho^2
	\right\rangle	
	\mp er \left\langle
		\hat \Theta \hat \lambda \hat \rho
	\right\rangle = 0 ,
$$
where we have introduced the following notation:
$$
	G_{\bar \Theta \rho}^\prime(\theta, \theta) =
	\left.
		\frac{d G_{\Theta(\theta) \rho(\theta')}}{ d \theta}
	\right|_{\theta' = \theta} .
$$
Here $\bar \Theta$ denotes that the derivative is taken with respect to the function $\Theta$ with the bar. The Green function
$G_{\Theta \rho}(\theta_1, \theta_2)$ is a function of two variables $\theta_1$ and $\theta_2$; hence, one has to have one more differential equation for the variable
 $\theta_2$. This equation can be obtained by multiplying the equation \eqref{NP_160} on the left by $\hat \Theta$ and by performing the quantum averaging,
$$
	\frac{1}{\sin \theta} \left(
		\sin \theta {G}^\prime_{\Theta \bar\rho}
	\right)^\prime = \pm er^{3/2}\frac{
		\left\langle \hat \Theta^3 \right\rangle
	}{4} .
$$
Similarly, one can obtain equations for the Green functions
$G_{\Theta \lambda}(\theta_1, \theta_2)$ and $G_{\Theta \Theta}(\theta_1, \theta_2)$.
As a result,
we arrive at an infinite set of equations for an infinite number of Green's functions
\begin{align}
	\left\langle \hat \Theta\right \rangle^\prime \mp
	\left\langle \hat \Theta \right\rangle
	\left[
		\cot \theta \left(\mp \frac{1}{2} + n \right) +
		\frac{m}{\sin \theta}
	\right] \mp er G_{\Theta \rho}
	\mp e r G_{\Theta \lambda} &= 0 ,
\label{NP_180}\\
	\frac{1}{\sin \theta} \left(
			\sin \theta \left\langle \hat \phi \right\rangle^\prime
		\right)^\prime &=
		- e r^{3/2} \frac{G_{\Theta \Theta}}{2}  ,
\label{NP_190}\\
	\frac{1}{\sin \theta} \left(
		\sin \theta \left\langle {\hat\rho}\right\rangle^\prime
	\right)^\prime &= \pm e r^{3/2} \frac{G_{\Theta \Theta}}{4}  ,
\label{NP_200}\\
	\left\langle \hat \lambda \right\rangle^\prime &=
	- \left\langle \hat \rho\right\rangle^\prime \cos \theta ,
\label{NP_210}\\
	G_{\bar \Theta \rho}^\prime \mp G_{\Theta \rho}
	\left[
		\cot \theta \left(\mp \frac{1}{2} + n \right) +
		\frac{m}{\sin \theta}
	\right]
	\mp er \left\langle
		\hat \Theta \hat \rho^2
	\right\rangle	
	\mp er \left\langle
		\hat \Theta \hat \lambda \hat \rho
	\right\rangle &= 0 ,
\label{NP_220}\\
	\frac{1}{\sin \theta} \left(
		\sin \theta {G}^\prime_{\Theta  \bar\rho}
	\right)^\prime &= \pm er^{3/2} \frac{
		\left\langle \hat \Theta^3 \right\rangle}{4},
\label{NP_230}\\
	G_{\bar \Theta \lambda}^\prime \mp G_{\Theta \lambda}
	\left[
		\cot \theta \left(\mp \frac{1}{2} + n \right) +
		\frac{m}{\sin \theta}
	\right]
	\mp er \left\langle
		\hat \Theta \hat \rho  \hat \lambda
	\right\rangle
	\mp er \left\langle
		\hat \Theta \hat \lambda^2
	\right\rangle	&= 0 ,
\label{NP_240}\\
	G_{\Theta \bar \lambda}' &= -
	G_{\Theta \bar \rho}' \cos \theta ,
\label{NP_245}\\
	G_{\bar \Theta \Theta}^\prime
	\mp G_{\Theta \Theta}
	\left[
		\cot \theta \left(\mp \frac{1}{2} + n \right) +
		\frac{m}{\sin \theta}
	\right]
	\mp er \left\langle
		\hat \Theta \hat \rho \hat \Theta
	\right\rangle
	\mp er \left\langle
		\hat \Theta \hat \lambda \hat \Theta
	\right\rangle	&= 0 .
\label{NP_250}\\
	\dots & = \dots
\label{NP_260}
\end{align}
As expected, Eqs.~\eqref{NP_220}, \eqref{NP_240}, and \eqref{NP_250} contain new 3-point Green's functions,
\begin{align}
\begin{split}
	 G_{\Theta \rho \rho}\left(
			\theta_1, \theta_2, \theta_3
		\right) &=
	\left\langle
		\hat \Theta(\theta_1) \hat \rho(\theta_2) \hat \rho(\theta_3)
	\right\rangle, \quad
	G_{\Theta \lambda \rho}\left(
			\theta_1, \theta_2, \theta_3
		\right) =
	\left\langle
		\hat \Theta(\theta_1) \hat \lambda(\theta_2) \hat \lambda(\theta_3)
	\right\rangle,
\\
		G_{\Theta \Theta \Theta}\left(
			\theta_1, \theta_2, \theta_3
		\right) &=
	\left\langle
		\hat \Theta(\theta_1) \hat \Theta(\theta_2) \hat \Theta(\theta_3)
	\right\rangle , \,\,\,\,
	G_{\Theta \rho \lambda}\left(
		\theta_1, \theta_2, \theta_3
	\right)  = \left\langle
		\hat \Theta(\theta_1) \hat \rho(\theta_2) \hat \lambda(\theta_3)
	\right\rangle ,
\\
		G_{\Theta \lambda \lambda}\left(
		\theta_1, \theta_2, \theta_3
	\right) &=
	\left\langle
		\hat \Theta(\theta_1) \hat \lambda(\theta_2) \hat \lambda(\theta_3)
	\right\rangle	, \quad
	G_{\Theta \rho \Theta}\left(
		\theta_1, \theta_2, \theta_3
	\right) =
	\left\langle
		\hat \Theta(\theta_1) \hat \rho(\theta_2) \hat \Theta(\theta_3)
	\right\rangle ,
\\
		G_{\Theta \lambda \Theta}\left(
		\theta_1, \theta_2, \theta_3
	\right) &=
	\left\langle
		\hat \Theta(\theta_1) \hat \lambda(\theta_2) \hat \Theta(\theta_3)
	\right\rangle ,
\nonumber
\end{split}
\end{align}
for which in turn one has to have differential equations determining these Green's functions. As a result of this process,
we finally get an infinite set of differential equations describing all Green's functions of the operators
$\hat \Theta, \hat \rho, \hat \phi,$ and $\hat \lambda$ appearing in the operator equations~\eqref{NP_140}-\eqref{NP_170}.

As was mentioned above, the infinite set of equations obtained can scarcely be solved explicitly, and hence the question of its approximate solving arises.
Using the experience accumulated in turbulence modeling, one can assume that this can be done by cutting off the infinite set of equations to a finite one using some
physical assumptions concerning higher-order Green's functions. In doing so, one can involve the following suppositions:
\begin{itemize}
\item One can neglect $n-$th order Green's functions compared with $(n-1)-$th order Green's functions.
\item $n-$th order Green's functions are polylinear combinations of lower-order Green's functions.
\item One can use the energy conservation law together with some physically reasonable propositions concerning its separate components.
\item And so on \ldots
\end{itemize}

\section{Discussion and conclusions}
\label{discussion}

We have considered the Dirac equation and Maxwell's electrodynamics in $\mathbb{R} \times S^3$ spacetime. The distinctive feature of these theories on the Hopf bundle is that they have discrete spectra of solutions both for the Dirac equation and for Maxwell's electrodynamics.
This is a consequence of the fact that a three-dimensional sphere $S^3$ is a compact space. For the Dirac equation,
this was shown explicitly by finding discrete solutions in analytic form. For the Maxwell equations, we have obtained numerical solutions,
as well as solutions for some particular cases, and the analysis of these solutions permits us to assume that a discrete spectrum of the solutions does exist.

The presence of the discrete spectrum allows one to quantize the free, noninteracting Dirac and Maxwell fields on the Hopf bundle. For the Dirac equation, the quantization is suggested by introducing the creation and annihilation operators for the corresponding quantum states. The standard anticommutation relations are imposed on these operators, and an additional numerical factor in the right-hand side of the anticommutator is introduced.
Using these relations, the propagator for the spinor field is calculated. The calculations indicate that this propagator is a sum over the discrete spectrum
numbered by the quantum numbers $m,n$ and $l$. In order to ensure the convergence of the sum, it is necessary that the aforementioned factor would possess a perfectly definite dependence on the quantum number $n, m , l$.

For the free electromagnetic field, a similar scheme of quantization has been suggested.

The most important part of the present study is the procedure of nonperturbative quantization suggested for the interacting Dirac and Maxwell fields.
Following Heisenberg~\cite{heis}, we have replaced the classical equations by equations for operators of the corresponding fields.
Since the operator equation can scarcely be solved somehow, it is replaced by an infinite set of equations for Green's functions.
Such a set is known as the Schwinger-Dyson equations, but they are usually employed in perturbative quantum field theory.

To illustrate the suggested scheme of nonperturbative quantization, we have considered some physical system possessing
perfectly definite $\mathfrak{Ansatze}$ for the spinor and electromagnetic fields. This gives the much more simple set of equations,
for which we have written out the first few equations for 1- and 2-point Green's functions.
For the equations describing 2-point Green's functions, we have explicitly written out 3-point Green's functions appearing in these equations.

The significance of examination of the scheme of nonperturbative quantization is that if, in nature, some fields are quantized, then apparently there
should exist a mathematically well-defined quantization procedure for \emph{any} fields, including those that are not quantized
due to the perturbative nonrenormalizibilty of the theory.

Let us note some features of the nonperturbative quantization.
\begin{itemize}
\item The properties of the operators of interacting fields can differ drastically from those of free fields.
For example, in the quantum theory of strongly interacting fields, there can exist static field configurations similar to those described by soliton,
monopole, instanton, etc. solutions in classical field theory.
\item The properties of the operators of interacting fields cannot be assigned by their commutators/anticommutators.
The algebra of these fields is much more complicated compared with the algebra given only by commutators/anticommutators.
These properties are determined by the infinite set of the Schwinger-Dyson equations as a whole.
\item For strongly interacting quantum fields, it is impossible to introduce creation and annihilation operators,
since the field operators cannot be represented as a superposition of plane waves with the coefficients that are creation and annihilation operators.
\item Separate consideration of the notion of quantum state is required, since in perturbative quantum field theory quantum states
are defined using creation and annihilation operators. According to what has been said in the previous item,
such operators cannot be defined for strongly interacting fields, and hence the definition of quantum states requires special consideration.
\item It is possible that the properties of the operators of interacting fields and of quantum states are related to the properties of the complete set of Green's functions
defined by the Schwinger-Dyson equations.
\end{itemize}

\section*{Acknowledgments}
We gratefully acknowledge
the Research Group Linkage Programme of the Alexander von Humboldt Foundation for the support of this research.
We are also grateful to Grant in Fundamental Research in Natural Sciences by the Ministry of Education and Science of the Republic of Kazakhstan.

\end{document}